\documentclass[11pt, a4paper]{article}

\makeatletter%

\usepackage{ifthen}

\newcommand{\MyMacro}[4]{
\provideboolean{#1}
\setboolean{#1}{#2}
\providecommand{#3}{\ifthenelse{\boolean{#1}}}
\providecommand{#4}{\ifthenelse{\not \boolean{#1}}}
}

\MyMacro{is_slides}{false}{\Ifslides}{\IfNotslides}

\MyMacro{use_packages}{true}{\IfUsepackages}{\IfNotUsepackages}

\MyMacro{plain_numbering}{true}{\IfPlainNumbering}{\IfNotPlainNumbering}

\MyMacro{single_numbering}{false}{\IfSingleNumbering}{\IfNotSingleNumbering}

\MyMacro{is_llncs}{false}{\Ifllncs}{\IfNotllncs}

\MyMacro{is_rus}{false}{\Ifrus}{\IfNotrus}



\makeatletter%

\newcommand{\ifndef}[2]{\ifthenelse{\isundefined{#1}}{#2}{}}

\newcommand{\mydef}[2]{\def#1{#2}}

\newcommand{\nospell}[1]{#1}  %

\makeatletter
\newcommand{\myusepackage}[2][]{\@ifpackageloaded{#2}{} %
{\ifthenelse{\equal{}{#1}} {\usepackage{#2}} {\usepackage[#1]{#2}} }}
\makeatother

\myusepackage{amssymb}
\myusepackage{amsmath}  %

\Ifllncs{ }{}
\myusepackage{amsthm}  %

\IfUsepackages{
\myusepackage[T1,OT2,T2A]{fontenc}
\DeclareTextSymbolDefault{\CYRYAT}{OT2}
\DeclareTextSymbolDefault{\cyryat}{OT2}
\DeclareTextSymbolDefault{\CYRFITA}{OT2}
\DeclareTextSymbolDefault{\cyrfita}{OT2}
\DeclareTextSymbolDefault{\CYRIZH}{OT2}
\DeclareTextSymbolDefault{\cyrizh}{OT2}
\myusepackage[utf8]{inputenc}
\let\f\relax

\Ifrus{
\usepackage[english,russian]{babel}
}{}

\myusepackage{latexsym}
\myusepackage{amsfonts}
\myusepackage{units}    %
\myusepackage{psfrag}   %
\myusepackage{url}   %
\myusepackage{hyperref}  %
\myusepackage{hyphenat}  %
\myusepackage{microtype}   %
\myusepackage{marginnote}  %
\Ifslides{
\myusepackage{mathrsfs}   %
\myusepackage{beamerthemesplit}
\usetheme[secheader]{Boadilla}
\usenavigationsymbolstemplate{}
}{  %
\myusepackage{enumitem}  %
\IfNotllncs{
\myusepackage[dvips]{graphicx}
\myusepackage[usenames,dvipsnames]{color}
}{
\myusepackage{graphicx}
}
}  %
}{} %

\Ifrus{
\newcommand{\dgDefinition}{Определение}
\newcommand{\dgFact}{Факт}
\newcommand{\dgQuestion}{Вопрос}
\newcommand{\dgLemma}{Лемма}
\newcommand{\dgCorollary}{Следствие}
\newcommand{\dgProposition}{Предложение}
\newcommand{\dgClaim}{Утверждение}
\newcommand{\dgTheorem}{Теорема}
\newcommand{\dgProblem}{Проблема}
\newcommand{\dgRemark}{Замечание}
\newcommand{\dgConjecture}{Предположение}
\newcommand{\dgResult}{Результат}
\newcommand{\dgProofOf}{\proofname}
}{
\newcommand{\dgDefinition}{Definition}
\newcommand{\dgDefinitions}{Definitions}
\newcommand{\dgFact}{Fact}
\newcommand{\dgFacts}{Facts}
\newcommand{\dgQuestion}{Question}
\newcommand{\dgQuestions}{Questions}
\newcommand{\dgLemma}{Lemma}
\newcommand{\dgLemmas}{Lemmas}
\newcommand{\dgCorollary}{Corollary}
\newcommand{\dgCorollaries}{Corollaries}
\newcommand{\dgProposition}{Proposition}
\newcommand{\dgPropositions}{Propositions}
\newcommand{\dgClaim}{Claim}
\newcommand{\dgClaims}{Claims}
\newcommand{\dgTheorem}{Theorem}
\newcommand{\dgTheorems}{Theorems}
\newcommand{\dgProblem}{Problem}
\newcommand{\dgProblems}{Problems}
\newcommand{\dgRemark}{Remark}
\newcommand{\dgRemarks}{Remarks}
\newcommand{\dgConjecture}{Conjecture}
\newcommand{\dgConjectures}{Conjectures}
\newcommand{\dgResult}{Result}

\newcommand{\dgProofOf}{\proofname\ of}
}

\IfPlainNumbering{
\ifndef{\theorem}{\newtheorem{theorem}{\dgTheorem}}
}{
\ifndef{\theorem}{\newtheorem{theorem}{\dgTheorem}[section]}
}

\Ifslides{}{  %
\IfSingleNumbering{
\ifndef{\lemma}{\newtheorem{lemma}[theorem]{\dgLemma}}
\ifndef{\corollary}{\newtheorem{corollary}[theorem]{\dgCorollary}}
\ifndef{\conjecture}{\newtheorem{conjecture}[theorem]{\dgConjecture}}
\ifndef{\remark}{\theoremstyle{remark} \newtheorem{remark}[theorem]{\dgRemark}}
\ifndef{\proposition}{\newtheorem{proposition}[theorem]{\dgProposition}}
\ifndef{\claim}{\newtheorem{claim}[theorem]{\dgClaim}}
\ifndef{\result}{\newtheorem{result}[theorem]{\dgResult}}
\ifndef{\problem}{\newtheorem{problem}[theorem]{\dgProblem}}
}{  %
\ifndef{\lemma}{\newtheorem{lemma}{\dgLemma}}
\ifndef{\corollary}{\newtheorem{corollary}{\dgCorollary}}
\ifndef{\conjecture}{\newtheorem{conjecture}{\dgConjecture}}
\ifndef{\remark}{\theoremstyle{remark} \newtheorem{remark}{\dgRemark}}
\ifndef{\proposition}{\newtheorem{proposition}{\dgProposition}}
\ifndef{\claim}{\newtheorem{claim}{\dgClaim}}
\ifndef{\result}{\newtheorem{result}{\dgResult}}
\ifndef{\problem}{\newtheorem{problem}{\dgProblem}}
}  %
}  %

\newtheoremstyle{mydefinition}  %
{\topsep}{\topsep}  %
{\slshape}  %
{}  %
{\bfseries}  %
{.}  %
{ }  %
{}  %

\newtheoremstyle{mynotation}  %
{\topsep}{\topsep}  %
{}  %
{}  %
{\bfseries\slshape}  %
{.}  %
{ }  %
{}  %

\newtheoremstyle{myremark}  %
{\topsep}{\topsep}  %
{\slshape}  %
{}  %
{\bfseries\slshape}  %
{:}  %
{ }  %
{}  %

\newtheoremstyle{myexample}  %
{\topsep}{\topsep}  %
{\itshape}  %
{}  %
{\slshape}  %
{:}  %
{ }  %
{\ul{\thmname{#1}}}  %

\newtheoremstyle{myclaims}  %
{\topsep}{\topsep}  %
{\slshape}  %
{}  %
{\bfseries\itshape}  %
{.}  %
{ }  %
{\thmname{#1}\thmnumber{ \!#2}\ifthenelse{\equal{}{#3}} %
{}{\textnormal{ \!(#3)}}}  %

\Ifslides{
\newtheoremstyle{remslide}{}{}{}{}{\itshape}{:}{ }{\thmname{#1}}
{\theoremstyle{remslide}\newtheorem{remark}{\dgRemark}}
\newtheorem{proposition}{\dgProposition}
\newtheorem{my_claim}{\dgClaim}
\newtheorem{result}{\dgResult}
}{ %
\ifndef{\notation}
{\theoremstyle{mynotation}\newtheorem*{notation}{Notation}}
\Ifllncs{

{\theoremstyle{myclaims}
\newtheorem{my_claim}[theorem]{\dgClaim}
\ifndef{\fact}{\newtheorem{fact}[theorem]{\dgFact}}
\ifndef{\question}{\newtheorem{question}[theorem]{\dgQuestion}}
\ifndef{\definition}{\newtheorem{definition}[theorem]{\dgDefinition}}
}
}{ %
{\theoremstyle{myremark}}
\IfSingleNumbering{
\ifndef{\definition}
{\theoremstyle{mydefinition}\newtheorem{definition}[theorem]{\dgDefinition}}
\ifndef{\example}
{\theoremstyle{myexample}\newtheorem{example}[theorem]{Example}}
{\theoremstyle{myclaims}
\newtheorem{my_claim}[theorem]{\dgClaim}
\ifndef{\fact}{\newtheorem{fact}[theorem]{\dgFact}}
\ifndef{\question}{\newtheorem{question}[theorem]{\dgQuestion}}
}
}{ %
\ifndef{\definition}
{\theoremstyle{mydefinition}\newtheorem{definition}{\dgDefinition}}
\ifndef{\example}
{\theoremstyle{myexample}\newtheorem{example}{Example}}
{\theoremstyle{myclaims}

\ifndef{\fact}{\newtheorem{fact}{\dgFact}}
\ifndef{\question}{\newtheorem{question}{\dgQuestion}}
}
} %
} %
} %

\IfNotllncs{
\newtheoremstyle{anystatement}{\topsep}{\topsep}{\itshape}{}{\bfseries}{.}{ }{\anystatementname}
{\theoremstyle{anystatement}}
\newcommand{\anystatementname}{}

}{}

\def\lf#1{\mathopen{}\left#1}
\def\rt#1{\right#1\mathclose{}}

\providecommand{\middle}{\big}
\newcommand{\md}{\middle}

\newcommand{\newident}[3][*]{\ifthenelse{\equal{*}{#1}}%
{\newcommand{#2}[1][]{\Ensuremath{\mathit{#3##1}}}}%
{\newcommand{#2}[1][]{\Ensuremath{\mathit{#3}}}}%
}

\newcommand{\newmat}[3][*]{\ifthenelse{\equal{*}{#1}}%
{\newcommand{#2}[1][]{\Ensuremath{#3##1}}}%
{\newcommand{#2}[1][]{\Ensuremath{#3}}}%
}

\newcommand{\providemat}[3][*]{\ifthenelse{\equal{*}{#1}}%
{\providecommand{#2}[1][]{\Ensuremath{#3##1}}}%
{\providecommand{#2}[1][]{\Ensuremath{#3}}}%
}

\newcommand{\newfunction}[2]{%
\newcommand{#1}[2][*]{\ifthenelse{\equal{*}{##1}}%
{\Ensuremath{#2\lf(##2\rt)}}%
{#2(##2)}}%
}

\newcommand{\MyMakeTheoMacros}[3]{
\newcommand{#2}[2][]{\ifthenelse{\equal{}{##1}}
{\begin{#1} ##2 \end{#1}}
{\begin{#1}\label{##1} ##2\end{#1}}}
\newcommand{#3}[3][]{\ifthenelse{\equal{}{##1}}
{\begin{#1}[\e{##2}] ##3 \end{#1}}
{\begin{#1}[\e{##2}]\label{##1} ##3\end{#1}}}
}

\makeatletter
\newtheorem*{rep@theorem}{\rep@title}
\newcommand{\newreptheorem}[2]{%
\newenvironment{rep#1}[1]{%
\def\rep@title{#2 \ref{##1}}%
\begin{rep@theorem}}%
{\end{rep@theorem}}}
\makeatother

\newcommand{\MyMakeDupTheoMacros}[7]{
\MyMakeTheoMacros{#1}{#2}{#3}
\newreptheorem{#1}{#6}
\newcommand{#4}[3]{
\newcommand{##2}{##3}
\begin{#1}\label{##1} ##2\end{#1}}
\newcommand{#5}[4]{
\newcommand{##2}{##4}
\begin{#1}{\e{##3}}\label{##1} ##2\end{#1}}
\newcommand{#7}[2]{\begin{rep#1}{##1} ##2 \end{rep#1}}
}

\Ifrus{
\newcommand{\MyMakeRefMacros}[3]
{\newcommand{#1}[2][]{\dg{Do not use referring macros!}}}
\newcommand{\MyMakeEqRefMacros}[3]
{\newcommand{#1}[2][]{\dg{Do not use referring macros!}}}
}{
\newcommand{\MyMakeRefMacros}[3]{\newcommand{#1}[2][]
{\ifthenelse{\equal{}{##1}}{#2~\ref{##2}}{#3~\ref{##1} and~\ref{##2}}}}

\newcommand{\MyMakeEqRefMacros}[3]{\newcommand{#1}[2][]
{\ifthenelse{\equal{}{##1}}{#2~\eqref{##2}}{#3~\eqref{##1} and~\eqref{##2}}}}
}

\Ifslides{
\newcommand{\bibentry}[9][*]{ \ifthenelse{\equal{*}{#1}}
{\bibitem[\nospell{#9}]{#2}{\textup #4}. \textrm{#5.} {\em #6, #7, #8}.}
{\bibitem[\nospell{#9}]{#2}{\textup #4}. {\em #5}.}
}

}{  %
\newcommand{\bibentry}[8]{
\Ifllncs
{\bibitem{#1} {\textup #3}.}
{\bibitem[\nospell{#8}]{#1} {\textup #3}.}
\ifthenelse{\equal{}{#6}}
{\newblock \textrm{#4.} \newblock{\em #5}, #7.}
{\newblock \textrm{#4.} \newblock{\em #5, #6}, #7.}
}

} %

\MyMakeTheoMacros{fact}{\fct}{\nfct}

\MyMakeRefMacros{\fctref}{\dgFact}{\dgFacts}

\MyMakeTheoMacros{question}{\quest}{\nquest}

\MyMakeRefMacros{\questref}{\dgQuestion}{\dgQuestions}

\MyMakeTheoMacros{notation}{\nota}{\nnota}

\Ifllncs{
\MyMakeDupTheoMacros{my_lemma}
{\lem}{\nlem}{\lemdup}{\nlemdup}{\dgLemma}{\lemrep}
}{ %
\MyMakeDupTheoMacros{lemma}
{\lem}{\nlem}{\lemdup}{\nlemdup}{\dgLemma}{\lemrep}
}

\MyMakeRefMacros{\lemref}{\dgLemma}{\dgLemmas}

\MyMakeDupTheoMacros{corollary}
{\crl}{\ncrl}{\crldup}{\ncrldup}{\dgCorollary}{\crlrep}

\MyMakeRefMacros{\crlref}{\dgCorollary}{\dgCorollaries}

\MyMakeTheoMacros{proposition}{\prp}{\nprp}

\IfNotllncs{
\newtheorem*{prp*}{\e{\dgProposition}}

}{}

\MyMakeRefMacros{\prpref}{\dgProposition}{\dgPropositions}

\MyMakeDupTheoMacros{my_claim}
{\clm}{\nclm}{\clmdup}{\nclmdup}{\dgClaim}{\clmrep}

\MyMakeRefMacros{\clmref}{\dgClaim}{\dgClaims}

\MyMakeDupTheoMacros{theorem}
{\theo}{\ntheo}{\theodup}{\ntheodup}{\dgTheorem}{\theorep}

\MyMakeRefMacros{\theoref}{\dgTheorem}{\dgTheorems}

\MyMakeTheoMacros{definition}{\defi}{\ndefi}

\MyMakeRefMacros{\defiref}{\dgDefinition}{\dgDefinitions}

\MyMakeTheoMacros{problem}{\prob}{\nprob}

\MyMakeRefMacros{\probref}{\dgProblem}{\dgProblems}

\MyMakeTheoMacros{remark}{\rem}{\nrem}

\MyMakeRefMacros{\remref}{\dgRemark}{\dgRemarks}

\MyMakeTheoMacros{conjecture}{\conj}{\nconj}

\MyMakeRefMacros{\conjref}{\dgConjecture}{\dgConjectures}

\renewcommand{\qedsymbol}{$\blacksquare$}

\newcommand{\prf}[2][]{\ifthenelse{\equal{}{#1}}%
{\begin{proof}\renewcommand{\qedsymbol}{$\blacksquare$}%
#2 \end{proof}}%
{\begin{proof}[\dgProofOf\ #1]%
\renewcommand{\qedsymbol}{$\blacksquare_{\mbox{\it{\scriptsize{#1}}}}$}%
#2 \end{proof}\renewcommand{\qedsymbol}{$\blacksquare$}}%
}

\newcommand{\prfstart}[1][]{\ifthenelse{\equal{}{#1}}%
{\begin{proof}\renewcommand{\qedsymbol}{$\blacksquare$}}%
{\begin{proof}[\dgProofOf\ #1]%
\renewcommand{\qedsymbol}{$\blacksquare_{\mbox{\it{\scriptsize{#1}}}}$}}%
}
\newcommand{\prfend}[1][*]{%
\ifthenelse{\equal{}{#1}}{\renewcommand{\qedsymbol}{$\blacksquare$}}{}%
\ifthenelse{\equal{*}{#1}}{}%
{\renewcommand{\qedsymbol}{$\blacksquare_{\mbox{\it{\scriptsize{#1}}}}$}}%
\end{proof}\renewcommand{\qedsymbol}{$\blacksquare$}%
}

\newcommand{\sect}[2][]{
\ifthenelse{\equal{*}{#2}}
{\section*}
{\ifthenelse{\equal{}{#1}}
{\section{#2}}
{\section{#2}\label{#1}}
}
}

\newcommand{\ssect}[2][]{
\ifthenelse{\equal{*}{#2}}
{\subsection*}
{\ifthenelse{\equal{}{#1}}
{\subsection{#2}}
{\subsection{#2}\label{#1}}
}
}

\MyMakeRefMacros{\chref}{Chapter}{Chapters}

\MyMakeRefMacros{\sref}{Section}{Sections}

\MyMakeRefMacros{\ssref}{Subsection}{Subsections}

\MyMakeRefMacros{\sssref}{Subsection}{Subsections}

\Ifslides{
\definecolor{DarkGreen}{rgb}{0,0.45,0.08}
\definecolor{LightBlue}{rgb}{0.122,0.016,0.855}

\renewcommand{\cite}[1]{[#1]}
}{} %

\MyMakeRefMacros{\figref}{Figure}{Figures}

\newcommand{\IfMathMode}[2]{\ifmmode{#1}\else{#2}\fi}

\newcommand{\Ensuremath}{\ensuremath}

\newcommand{\fbr}[1]{\IfMathMode%
{#1}{$#1$}}                     %

\newcommand{\fnbr}[1]{\mbox{\fbr{#1}}}  %

\newcommand{\fla}[2][*]{\ifthenelse{\equal{}{#1}}{\fbr{#2}}{\fnbr{#2}}}
\newcommand{\f}{\fla}

\newcommand{\malabel}[1]{\addtocounter{equation}{1}\tag{\theequation}\label{#1}}
\newcommand{\mal}[2][]{%
\ifthenelse{\equal{}{#1}}%
{\begin{align*} #2 \end{align*}}%
{\ifthenelse{\equal{P}{#1}}%
{\allowdisplaybreaks\begin{align*} #2%
\end{align*}\interdisplaylinepenalty=10000}%
{\begin{align*} \malabel{#1} #2 \end{align*}}%
}%
}

\newcommand{\m}{\mal}

\newcommand{\mac}{\substack}

\MyMakeEqRefMacros{\equref}{Equation}{Equations}

\MyMakeEqRefMacros{\expref}{Expression}{Expressions}

\MyMakeEqRefMacros{\inequref}{Inequality}{Inequalities}

\newcommand{\bref}[1]{(\ref{#1})}

\newcommand{\twocase}[4]%
{\begin{cases} #1 &\txt{#2}\\ #3 &\txt{#4}\end{cases}}

\newcommand{\chs}{\genfrac(){0cm}{}}  %

\providecommand{\E}[2][]{\mathop{\mathbf{E}}_{#1}\lf[{#2}\rt]}

\newcommand{\PR}[2][]{\mathop{\mathbf{Pr}}_{#1}\lf[{#2}\rt]}
\newcommand{\PRr}[3][]{\mathop{\mathbf{Pr}}_{#1}\lf[{#2}\vphantom{|_1^1}\md|\vphantom{|_1^1}{#3}\rt]}

\renewcommand{\U}[1][]{\ifthenelse{\equal{}{#1}}%
{{\cal U}}%
{{\cal U}_{#1}}}

\newcommand{\pss}[1][]{\nospell{\ifthenelse{\equal{}{#1}}%
{\txt{'s}}%
{\fla{#1\txt{'s}}}}}

\newcommand{\pl}[1][]{\nospell{\ifthenelse{\equal{}{#1}}%
{\mskip-6mu\stackrel{\text-}{}\mskip-4mu\txt{s}}%
{\fla{#1\mskip-6mu\stackrel{\text-}{}\mskip-4mu\txt{s}}}}}

\newcommand{\ord}[1][]{\nospell{\ifthenelse{\equal{}{#1}}%
{\txt{'th}}%
{\ifthenelse{\equal{1}{#1}}{$1\txt{'st}$}{\ifthenelse{\equal{2}{#1}}{$2\txt{'nd}$}{\ifthenelse{\equal{3}{#1}}{$3\txt{'rd}$}{\fla{#1\txt{'th}}}}}}}}

\newcommand{\fr}[3][*]{%
\ifthenelse{\equal{*}{#1}}%
{\frac{#2}{#3}}{}%
\ifthenelse{\equal{/}{#1}}%
{\nicefrac{#2}{#3}}{}%
\ifthenelse{\equal{}{#1}}%
{\lf.#2\md/#3\rt.}{}%
\ifthenelse{\equal{p_}{#1}}%
{\lf.\lf(#2\rt)\md/#3\rt.}{}%
\ifthenelse{\equal{_p}{#1}}%
{\lf.#2\md/\lf(#3\rt)\rt.}{}%
\ifthenelse{\equal{pp}{#1}}%
{\lf.\lf(#2\rt)\md/\lf(#3\rt)\rt.}{}%
}

\newcommand{\dr}{\nicefrac}

\newcommand{\sq}{\sqrt}

\newcommand{\set}[2][]{\ifthenelse{\equal{}{#1}}%
{\Ensuremath{\lf\{#2\rt\}}}%
{\Ensuremath{\lf\{#2\vphantom{|_1^1}\md|\vphantom{|_1^1}#1\rt\}}}}

\newcommand{\Min}[2][]{\ifthenelse{\equal{}{#1}}%
{\Ensuremath{\min\lf\{#2\rt\}}}%
{\Ensuremath{\min\lf\{#2\vphantom{|_1^1}\md|\vphantom{|_1^1}#1\rt\}}}}

\newcommand{\Minn}[3][]{\ifthenelse{\equal{}{#1}}%
{\Ensuremath{\min_{#2}\lf\{#3\rt\}}}%
{\Ensuremath{\min_{#2}\lf\{#3\vphantom{|_1^1}\md|\vphantom{|_1^1}#1\rt\}}}}

\newcommand{\Maxx}[3][]{\ifthenelse{\equal{}{#1}}%
{\Ensuremath{\max_{#2}\lf\{#3\rt\}}}%
{\Ensuremath{\max_{#2}\lf\{#3\vphantom{|_1^1}\md|\vphantom{|_1^1}#1\rt\}}}}

\newfunction{\asO}{O}
\newfunction{\astO}{\tilde O}
\newfunction{\aso}{o}
\newfunction{\asOm}{\Omega}
\newfunction{\astOm}{\tilde \Omega}
\newfunction{\asom}{\omega}
\newfunction{\asT}{\Theta}

\mydef{\01}{\set{0,1}}

\newcommand{\sz}[2][]{\ifthenelse{\equal{}{#1}}%
{\Ensuremath{\lf|#2\rt|}}%
{\Ensuremath{\lf|#2\rt|_{#1}}}}

\newcommand{\ssz}[2][]{\lf|\set[{#1}]{#2}\rt|}

\newcommand{\txt}[1]{\textrm{#1}}  %

\newcommand{\Cl}{\mathcal}  %

\DeclareMathAlphabet{\mathlowcal}{OT1}{pzc}{m}{it}
\newcommand{\Cll}{\mathlowcal}

\newident[]{\N}{\mathcal N_{#1}}
\newident[]{\R}{\mathcal R_{#1}}

\newident{\NP}{NP}
\newident{\AM}{AM}

\newident{\Disj}{Disj}

\mydef{\l(}{\lf(}
\mydef{\r)}{\rt)}

\newcommand{\nin}{\not\in}  %

\newcommand{\Then}{\Longrightarrow}

\newcommand{\dt}{\cdot}
\newcommand{\tm}{\cdot}
\newcommand{\sbseq}{\subseteq}
\newcommand{\sbs}{\subset}
\newcommand{\deq}{\stackrel{\textrm{def}}{=}}
\mydef{\=}{\equiv}
\newcommand{\unin}{\mathrel{\subset\mkern-13.1mu\sim}}  %

\providemat{\NN}{\mathbb{N}}

\newcommand{\ds}[1][]
{\ifthenelse{\equal{}{#1}}{\allowbreak\dots}{#1\allowbreak\dots#1}}
\newmat{\dc}{\ds[,]}

\newcommand{\abstart}{\begin{abstract}}
\newcommand{\abend}{\end{abstract}}

\newcommand{\quo}[1]{\begin{quote} #1 \end{quote}}

\Ifslides{
\newcommand{\itemi}[2][]{\ifthenelse{\equal{}{#1}}%
{\begin{itemize} #2 \end{itemize}}%
{\begin{itemize}[#1] #2 \end{itemize}}}

}{ %
\newcommand{\itemi}[2][*]{\ifthenelse{\equal{*}{#1}}%
{\begin{itemize}[noitemsep] #2 \end{itemize}}%
{\begin{itemize}[#1] #2 \end{itemize}}}

}  %

\protected \def \dg #1{%
\textcolor{Red}
{
{\normalmarginpar\marginnote{\bl{DG's comment}}}
{\reversemarginpar\marginnote{\bl{DG's comment}}\\}
\IfMathMode{
~~~\txt{#1}~
}{
~\\~~~#1~\\
{\normalmarginpar\marginnote{\bl{\ul{------}}}}
{\reversemarginpar\marginnote{\bl{\ul{------}}}\\}
}
}
\ClassWarning{My Macros}{#1}
}

\newcommand{\fn}[2][]{%
\IfMathMode{}{}%
\ifthenelse{\equal{}{#1}}%
{\footnote{#2}}%
{\footnote{\label{#1}#2}}%
}


\Ifslides{
\newcommand<>{\e}[1]{\IfMathMode%
{\only#2{\textcolor{BrickRed}}{#1}}%
{\emph{\only#2{\textcolor{BrickRed}}{#1}}}%
}
}{ %
\newcommand{\e}{\emph}
}  %

\newcommand{\bl}[1]{{\bf #1}} %

\providecommand{\ul}[1]{\underline{#1}} %

\makeatother%

\IfNotllncs{

\setlength{\textheight}{9.25in}
\setlength{\topmargin}{-.55in}
\setlength{\textwidth}{6.5in}
\setlength{\evensidemargin}{-.1in}
\setlength{\oddsidemargin}{-.1in}

\bibliographystyle{alpha}
}{}

\makeatother%

\newident{\II}{\Cl{II}}
\newident{\IIAB}{\Cl{II}_{A,B}}

\title{The communication complexity\\
of the inevitable intersection problem}

\newcommand{\instDG}{Institute of Mathematics, Academy of Sciences, \v Zitna 25, Praha 1, Czech Republic.}

\newcommand{\thanksDG}{Partially funded by the grant 19-27871X of GA \v CR.
Part of this work was done while visiting the Centre for Quantum Technologies at the National University of Singapore, and was partially supported by the Singapore National Research Foundation, the Prime Minister's Office and the Ministry of Education under the Research Centres of Excellence programme under grant R 710-000-012-135.}

\author{Dmitry Gavinsky\thanks{\instDG\ \thanksDG}
}

\begin{document}

\maketitle

\thispagestyle{empty}

\abstart

\e{Set disjointness} (\Disj) is a central problem in communication complexity.
Here Alice and Bob each receive a subset of an \f n-element universe, and they need to decide whether their inputs intersect or not.
The communication complexity of this problem is relatively well understood, and in most models, including -- most famously -- \e{interactive randomised communication with bounded error} (\R), the problem requires much communication.

In this work we were looking for a variation of \Disj, as natural and simple as possible, for which the known lower bound methods would fail, and thus a new approach would be required in order to understand its \R-complexity.
The problem that we have found is a \e{relational} one:\ each player receives a subset as input, and the goal is to find an element that belongs to both players.
We call it \e{inevitable intersection} (\II).

The following list of its properties seem to let \II\ resist the old lower bound techniques:\itemi{
\item the domain of \II\ is $A\times B$, the product of the players' individual input spaces;
\item $A\times B$ only contains intersecting pairs of subset;
\item the input comes from the uniform distribution over $A\times B$;
\item $A\times B$ is chosen in a randomised fashion, both $A$ and $B$ being uniformly-random subsets of $2^{[n]}$ of size $2^{n^{\asT1}}$.
}
In particular, complexity analysis of \II\ cannot be based on the hardness of \Disj\ (as no pair in $A\times B$ is disjoint); moreover, it cannot be based on any argument based on discrepancy (including corruption, smooth discrepancy and the like), as the problem allows for a cover of $A\times B$ by $n$ perfectly-monochromatic rectangles.

We are using an ad hoc technique to show that \II\ is \e{ultimately hard}:\ it requires $\asOm{\log\sz A}$ bits of interactive randomised communication.
Besides its ability -- apparently unique -- to capture the complexity of the inevitable intersection, the new technique can also be applied to other ``search-like'' problems (including \Disj), thus providing new insight into their communicational hardness.
\abend

\newpage

\setcounter{page}{1}

\sect[s_intro]{Introduction}

Unstructured search is a basic computational paradigm and its natural instance in the context of communication complexity is the \e{intersection problem}.
Here each player receives a subset of $[n]$ and the goal is \e{to decide whether the intersection of these subsets is empty} in the \e{decision version}, and \e{to find an element from the intersection when it is not empty} in the \e{search version} of the problem.
The decision version is exactly the same as the well-known \e{set disjointness problem} (\Disj).
The search version looks more demanding, but for most communication models the decision and the search version are, essentially, equivalent, as the search version can be reduced to a short series of decision instances (e.g., via binary search).

The most common in the context of communication complexity is, probably, the model of \e{randomised interactive communication}:\ there are two players, Alice and Bob, who are allowed to use random bits and exchange messages in order to find an answer with respect to the input that is divided between the two of them.
The complexity of a given protocol is defined as the maximum possible number of bits exchanged before producing the answer, when each player's input has length at most $n$.
The answer must be correct with non-trivial constant probability -- say, at least $p$ when the input distribution is assumed to be such that no single answer can be correct with probability more than $p-\asOm1$.
We denote this model by \R.
The complexity of a problem $P$ in \R\ is the minimum complexity of a valid protocol for solving $P$, and we write $\R(P)$ to denote it.

No efficient solution for \Disj\ exists in most communication models, so understanding its complexity amounts to very little more than proving a tight lower bound.
In particular,~\cite{KS87_The_Pr} has shown that $\R(\Disj)\in\asOm n$ (see also~\cite{BFS86_Co_Cl, R92_On_the}).

Virtually all known lower bound methods for \Disj\ are based on the \e{hardness of witnessing non-intersection of the input sets}:
Informally, while it is easy to demonstrate that the given subsets overlap (e.g., by pointing to an element from the intersection), it can be rather difficult to certify that the given subsets are disjoint.\fn
{
This can be formally expressed in terms of \e{non-deterministic} communication complexity \N, which captures the complexity of \e{certifying satisfying instances} of the target Boolean function (viewed as a predicate).
The \N-complexity is \asT{n} for disjointness and \asT{\log n} for non-disjointness.
}
Due to the fact that a communication transcript can be used as a correctness certificate of the answer produced by a valid protocol, hardness of witnessing non-intersection implies a lower bound on the communication complexity of \Disj\ and a number of its modifications.

Given the importance and the natural appeal of both the problem \Disj\ and the model \R, in this work we have been looking for as simple as possible variation of the former, whose complexity analysis in the latter would lead to a new approach:
While the known ones undoubtedly provide deep insight on the core of hardness of \Disj, there seems to remain something mysterious about it, as witnessed, for example, by the fact that we know nothing about the complexity of \Disj\ in the model of \AM\ in communication complexity, which allows both randomness (like \R) and non-determinism (like \N, the ``communication complexity \NP'').\fn
{
The bound $\N(\Disj)\in\asOm n$ follows, essentially, from the same argument as the one used to establish $\R(\Disj)\in\asOm n$.
However, the same approach (as well as all others that we are aware of) falls astonishingly short of providing any insight on the complexity of \Disj\ in \AM, a model that can be viewed as the natural closure, where the strength of both $\N$ and $\R$ is present.
Resolving the communication complexity of \Disj\ in \AM\ is one of the most important open problems in the field.
}

A communication problem that we call \e{inevitable intersection} (\II) is an instance of the intersection search problem:\ both Alice and Bob receive a subset as input, and their goal is to find an element that belongs to both players.
Formally, \IIAB\ is the following problem:
\quo{
Let $A,B\sbseq\01^n$ be such that
\m{\forall a\in A, b\in B:~a\cap b\neq\emptyset,}
where a binary string is viewed as the set of its coordinates with value ``$1$''.
Alice receives $a\in A$, Bob receives $b\in B$ and they have to output some $i\in a\cap b$.
}
We will analyse the (expected) \R-complexity of \IIAB\ when $A$ and $B$ are uniformly-random subsets of $\set[\sz{x}=n^{\dr35}]{x\in\01^n}$ of size $2^{\dr{\sq[5]n}5}$.~\fn
{
The constants in the definition have been chosen to allow as simple as possible tight lower-bound analysis, based on the new technique (cf.~\crlref{c_II_low}).
The approach seems to be adjustable to give tight results for any possible choice of parameters (though we haven't verified the details), but the required changes are rather technical and do not seem to provide enough additional insight to justify the more involved argument.
}

The definition of \II\ is analogous to the well-known monotone case of Karchmer-Wigderson games~\cite{KW88_Mon,RW92_Mon}, where lower bounds are known for some instances.
Those techniques don't seem applicable to the randomised case of \IIAB, as defined above.
The special properties of \IIAB\ that, as we believe, have allowed it to frustrate the previous lower-bound methods are the following:\itemi{
\item the input space $A\times B$ has a product structure, and still it only contains intersecting pairs of subsets;
\item $A\times B$ is chosen in a randomised fashion, both $A$ and $B$ being uniformly-random subsets of $2^{[n]}$ of the required size.
}

As mentioned above, virtually all known hardness arguments for \Disj\ are based on the observation that it is hard for a communication protocol to certify that the input subsets are disjoint.
More formally, an efficient communication protocol in any of those communication models where \Disj\ is known to be hard necessarily implies existence of a relatively large nearly-monochromatic (with respect to the target problem) combinatorial rectangle in the space of pairs of input values; accordingly, showing that no such rectangle is possible proves hardness of the problem for these models.
In the case of \Disj\ there exist large rectangles that are biased towards intersecting input pairs; however, it has been shown that no large rectangle can be sufficiently biased towards non-intersecting pairs (which is the formal way to say that being disjoint is hard to witness), which implies that \Disj\ has no efficient protocol.

In the case of \IIAB, on the other hand, not only do large nearly-monochromatic rectangles exist, but in fact the whole input space $A\times B$ can be covered by $n$ perfectly-monochromatic rectangles:\ e.g., for any $i\in[n]$ the set of all pairs intersecting on $i$ form a rectangle, and the union of all such rectangles equals $A\times B$.
This fact alone makes it impossible to deduce the hardness of \II\ from an argument that is based on discrepancy (that is, from reasoning about the possibility of sufficient bias in large rectangles).

The \e{product structure} of the support $A\times B$ prevents the possibility to ``secretly'' add to it disjoint pairs and to base a lower bound argument on the contradiction between, on the one hand, the possibility to use a good \II-protocol for detecting disjoint pairs (where the protocol would not find a common element) and, on the other hand, the hardness of deciding whether the input pair is disjoint (to be shown via standard techniques).
Adding a non-intersecting pair to $A\times B$ would require extending the marginal support of at least one of the players; even if the input encoding scheme would allowing such an extension, a very short protocol could easily detect an input pair outside $A\times B$ and produce a special response (say, declare an error), thus frustrating the intended hardness reduction.
At the same time, the \e{lack of structure} in the families $A$ and $B$ themselves seems to make it impossible to embed hard instances of \Disj\ into instances of \IIAB.

Our way of making the complexity of \Disj\ difficult to analyse has one aspect in common with the problem of analysing the \AM-complexity of \Disj:
Namely, both questions escape the reach of the current techniques, partially, due to the fact that \e{a hardness statement cannot follow from reasoning about the discrepancy of large rectangles}.
In the case of \AM\ the model is such that an efficient protocol does not seem to guarantee existence of a large biased rectangle even in the case of total functions, and for promise functions or relations an efficient \AM\ protocol is known~\cite{K11_On_A} to be possible in some cases, where every rectangle is either exponentially small, or has error inverse-exponentially close to $\dr12$.
In our case the problem \II\ (obviously, a variation on \Disj) has been defined in such a way that it admits very large monochromatic rectangles, a few of which are enough to cover the whole input space.

We prove that \IIAB, as defined above, is \e{ultimately hard}:\ it requires $\asOm{\sq[5]n}=\asOm{\log\sz A}$ bits of interactive randomised communication (cf.~\crlref{c_II_low}).
We are not aware of a previously-known technique that would give a non-trivial lower bound for \IIAB.
Besides being able to capture the complexity of \IIAB\ in \R, the technique proposed in this work provides an alternative lower bound proof for many other cases of ``search-like'' problems (including the set disjointness), thus giving new insight on the aspects of their combinatorial structure that are responsible for their communication hardness.

\sect{Preliminaries}

We will write $[n]$ to denote the set $\set{1\dc n}\sbs\NN$.
Let $x(i)$ address the \ord[i] bit of $x$ for $x\in\01^n$ and $i\in{[n]}$.
Similarly, for $S\sbseq{[n]}$, let $x(S)$ denote the $\sz S$-bit string, consisting of (naturally-ordered) bits of $x$, whose indices are in $S$.

Let $\sz x$ denote the Hamming weight of $x\in\01^n$.
At times we will assume the trivial isomorphism between the \f n-bit strings and the subsets of $[n]$.
In particular, the notation $\chs{[n]}k$ will stand for $\set[\sz{x}=k]{x\in\01^n}$, and $x\cap y$ will address the set $\set[x_i=y_i=1]{i\in[n]}$.

For a discrete set $A$, we denote by $\U[A]$ the uniform distribution over its elements.
Sometimes (e.g., in subscripts) we will write ``$X\unin A$'' instead of ``$X\sim\U[A]$''.

We let $\log$ denote the base-$2$ logarithm; at times, we will write $\exp(\dt)$ instead of $e^{\dt}$ to avoid superscript congestion.

We will use the Chernoff bound in the following form (cf.~\cite{DM05_Con_Bo}).

\nclm[c_Che]{Chernoff bound}{Let $X_1\dc X_n$ be mutually independent random variables taking values in $[0,1]$ and $\E{X_i}\equiv\mu$.
Then for any $\Delta>0$,
\m{\PR{\fr1n\tm\sum_{i=1}^nX_i\ge\mu+\Delta}\le
e^{\fr{-n\Delta^2}{2\mu+\Delta}}}
and
\m{\PR{\fr1n\tm\sum_{i=1}^nX_i\le\mu-\Delta}\le
e^{\fr{-n\Delta^2}{2\mu}}.}
}

The following tail bound can be viewed as a variation on Markov's inequality.
\lem[l_mav]{Let $X$ be a random variable taking values in $[a,b]$, then for any $\Delta>0$,
\m{\PR{X<\E X+\Delta},~\PR{X>\E X-\Delta} ~\ge~ \Min{\fr\Delta{b-a},1}.}
}

\prf{Let $\lambda\deq\PR{X<\E X+\Delta}$, then
\m{\E X\ge(\E X+\Delta)\tm(1-\lambda)+a\tm\lambda,}
and therefore,
\m{\lambda\ge\fr\Delta{(\E X+\Delta)-a}.}
If $\E X+\Delta\le b$, then $\lambda\ge\fr\Delta{b-a}$; otherwise, $\lambda=1$ trivially.

The case of $\PR{X>\E X-\Delta}$ is similar.}

\ssect*{Communication complexity}

We will write $\R_p$ to denote the model of randomised interactive communication with worst-case probability of correct answer at least $p$.
When $p$ is obvious from the context or irrelevant, we drop the subscript $_p$.
We will denote by $\R_{\mu,p}$ the distributional version of $\R_p$, where the input distribution is assumed to be $\mu$.
For a communication problem $\Cl S$, we will write $\R(\Cl S)$, $\R_p(\Cl S)$ or $\R_{\mu,p}(\Cl S)$ to denote its complexity in the corresponding model.

One of the most studied communication complexity problems is set disjointness: Alice receives $x$ and Bob receives $y$ as input, and they have to decide whether the two sets overlap (note that this is a function:\ for every input pair, there is exactly one correct answer).

\ndefi[d_Disj]{Set disjointness problem, \Disj}
{For $x,y\sbseq[n]$, let
\m{\Disj(x,y)\deq\twocase{1}{if $x\cap y=\emptyset$}{0}{otherwise}.}
}

In this work we study the following problem.

\ndefi[d_II]{Inevitable intersection problem, \IIAB}
{For $A,B\sbseq\01^n$ such that
\m{\forall a\in A, b\in B:~a\cap b\neq\emptyset,}
let
\m{\IIAB\deq\set[i\in a\cap b]{(a,b,i)\in A\times B\times[n]}.}
}

Informally, when Alice receives $a$ and Bob receives $b$ as their input to \IIAB, a correct answer is any $i\in a\cap b$ (a correct answer does not have to be unique, so this is a relational problem).
Note that \IIAB\ can be viewed as a search version of \Disj\ with an additional constraint that $\forall a\in A, b\in B:a\cap b\neq\emptyset$.
The most important for us is the \e{syntactic nature} of this constraint:\ an instance of \IIAB\ is defined by the choice of $A,B\sbseq\01^n$ for every $n\in\NN$, and only those instances are valid where ``$a\cap b\neq\emptyset$'' is a \e{tautology}.

\sect[s_method]{Our argument}

A \e{deterministic} \f2-party communication protocol of length $c$ defines a partition of the input matrix into at most $2^c$ same-answer rectangles (the protocol is able to distinguish only between input pairs coming from different rectangles).
At the same time, if there exists an efficient \e{randomised} protocol, then there must exist a reasonably-accurate deterministic protocol for every input distribution $\mu$.
A typical rectangle defined by such a protocol must be relatively large (otherwise the union of all rectangles would be too small to cover the whole input matrix) and nearly-monochromatic (otherwise the protocol would not be sufficiently accurate).

In the case of \Disj\ one can find an input distribution $\mu$ such that no large (with respect to \f\mu) rectangle would consist mostly of non-intersecting input pairs, and at the same time, the probability of a pair of sets $(X,Y)\sim\mu$ to not intersect would be close to $\dr12$.
The above reasoning implies that if a short randomised protocol for \Disj\ were possible, the non-intersecting input pairs that are often produced by $\mu$ would have ``no rectangle to go'', thus contradicting the assumption and leading to the desired lower bound on the randomised communication complexity of \Disj.

In the case of \IIAB, non-emptiness of $a\cap b$ holds for every possible input pair $(a,b)\in A\times B$, so one cannot meaningfully ask ``\e{Where do non-intersecting input pairs go?}'':
If $a\nin A$ or $b\nin B$, at least one of the players would immediately notice the promise violation.

To analyse the communication complexity of \IIAB, we will use the following approach.
Consider a deterministic protocol of complexity $c$ that solves \IIAB\ with respect to the uniform (over $A\times B$) input distribution $\U$ with error at most $\dr12$.
This protocol corresponds to a partition of $A\times B$ into at most $2^c$ rectangles that are labelled by the answers of the protocol, such that $(X,Y)\sim\U$ belongs to a rectangle labelled by some $i\in X\cap Y$ with probability at least $\dr12$.

We would like to get a lower bound of the form $c\in n^{\asOm1}$; that is, we want to show that a \e{partition} of $A\times B$ with properties as described above must have size $\exp\l(n^{\asOm1}\r)$.
Note that there always exists a \e{cover} of $A\times B$ by $n$ perfectly-monochromatic rectangles:
\m[m_ri]{r_i\deq\set[a(i)=b(i)=1]{(a,b)\in A\times B}
,}
where the label of $r_i$ is ``$i$''.
So, we are looking for a property of large rectangles that would obstruct combining them into a partition of $A\times B$, but not into a cover of it.

Let us consider a partition $R$ of $A\times B$ into rectangles.
For all $i\in[n]$ and $r\in R$, let
\m{p(r,i)\deq\PRr[(X,Y)\sim\U]{X(i)=Y(i)=1}{(X,Y)\in r}
.}
For a typical $r_0\in R$ that is labelled by ``$i_0$'', we expect $p(r_0,i_0)$ to be high, but what about the rest of \pl[i]?
We will see (cf.~\lemref{l_rec}) that if $r_0$ is large enough, then, informally speaking, $p(r_0,i)$ cannot be too different from the global (unconditional) expectation
\m{p_i\deq\PR[\U]{X(i)=Y(i)=1}}
for too many values of $i\in[n]$ (this is the case for the rectangles defined in \bref{m_ri}, for instance).
Intuitively, a rectangle pays in terms of the entropy of its uniformly-random element for making some of its bits biased (i.e., making $p(r_0,i)$ significantly different from $p_i$).

This property of large rectangles is enough to prove a strong lower bound on the cardinality of $R$ (cf.~\theoref{t_II_low}).
To understand how, let us consider the following extreme situation:\ the rectangles in $R$ are either \e{small} or \e{large}, and for a large $r_0\in R$ labelled by $i_0$ it holds that
\m{p(r_0,i)\approx\twocase
{\fr12}{if $i=i_0$}
{p_i}{otherwise}	
.}
Let us also assume $p_i\=\l(\fr kn\r)^2$ for some $k\ll n$ (this is almost true almost always if $A$ and $B$ are sufficiently large random subsets of $\chs{[n]}k$, which will be the case of interest to us).
If we assume by contradiction that $R$ is small, then a significant fraction of its rectangles must be large; let us again take the extreme case and assume that all \pss[R] rectangles are large.
Let $i_1\in[n]$ be such that at least $\dr1n$-fraction of $A\times B$ belongs to ``$i_1$''-labelled rectangles from $R$, and let us use the above assumptions to estimate $p_{i_1}$:
On the one hand, with some probability $q\ge\dr1n$, a uniformly-random input $(X,Y)$ belongs to a rectangle $r\in R$ labelled by ``$i_1$''; by our assumptions, this event contributes roughly $p(r,i_1)\tm q=\fr12\tm q$ to the probability that $X(i)=Y(i)=1$.
On the other hand, with probability $1-q$ the input belongs to a rectangle labelled differently, and we have assumed that in that case $X(i_1)=Y(i_1)=1$ with probability roughly $p_{i_1}$, so this event contributes about $p_{i_1}\tm(1-q)$.
Therefore,
\m{p_{i_1} \approx \fr12\tm q + p_{i_1}\tm(1-q)
~\Then~
p_{i_1} \approx \fr12
,}
which contradicts our assumption that $p_i\=\l(\fr kn\r)^2\ll1$.

Our argument can be summarised like this:
On the one hand, a large nearly-monochromatic rectangle in $R$ causes a noticeable deviation (increase) of the global probability that its label coordinate belongs to $X\cap Y$; on the other, large rectangles cannot efficiently absorb the deviations caused by other large rectangles; therefore, there must be many small rectangles in $R$, and the partition itself must be large.
While in order to prove that \Disj\ is communicationally hard one usually argues that in a short protocol \e{non-intersecting input pairs would have ``nowhere to go''}, we will argue that in a hypothetical short protocol for \IIAB\ there would be \e{excessive probabilities} of those intersection coordinates that the protocol typically outputs.

\sect[s_bound]{The communication complexity of \II}

We start by proving a lemma that limits witnessing against coordinate-wise intersections by a large input rectangle.

\lem[l_rec]{Let $1\le k<\fr n2$ and $1\le M'\le M\le\fr12\chs nk^{\dr12}$, such that
\m{\log\l(\fr M{M'}\r) \le \fr{\log M}3-5\log n.}
Then for
\m{\Delta=51\tm\fr{k^{\dr32}}n\tm\sq{\log\l(\fr M{M'}\r)+\log n}}
it holds that
\m{\Maxx{\mac{A'\sbseq A\\B'\sbseq B\\\sz{A'},\sz{B'}\ge M'\\T\sbseq[n]}}
{\sum_{i\in T}\l(\l(\fr kn\r)^2-\PR[(X,Y)\unin A'\times B']{X(i)=Y(i)=1}\r)}
<\Delta
}
with probability higher than $1-\exp(n-M^{\dr13})$ when $A$ and $B$ are uniformly-random subsets of $\chs{[n]}k$ of size $M$.
}

Informally, the lemma states that almost always with respect to $A$ and $B$, membership of the input pair $(X,Y)$ in a large rectangle $A'\times B'\sbseq A\times B$ cannot significantly decrease the probability that $X(i)=Y(i)=1$ for many $i\in[n]$; note that this probability equals $\l(\fr kn\r)^2$ when $X,Y\unin\chs{[n]}k$.
This lemma will be the core technical tool of the lower bound proof for \IIAB.

\prfstart[\lemref{l_rec}]
Consider some $A'\sbseq A\sbseq\chs{[n]}k$ and let $p_i\deq\PR[X\unin A']{X(i)=1}$, $\alpha\deq\sum_{i=1}^n\sz{p_i-\fr kn}$ and
\f{S\deq\set[p_i<\fr kn]{i\in[n]}.}
Let us see that if $\alpha$ is big enough, then $A'$ contains a non-negligible fraction of bit strings, whose projection to $S$ has unnaturally low Hamming weight.
As $\sum p_i=k$ by assumption, $\sum_{i\in S}\l(\fr kn-p_i\r)=\fr\alpha2$ and
\m{\E[X\unin A']{\sz{X(S)}}=\fr{k\tm\sz S}n-\fr\alpha2.}
Therefore by \lemref{l_mav},
\m{\PR[X\unin A']{\sz{X(S)}\le\fr{k\tm\sz S}n-\fr\alpha4}
\ge\fr[_p]{\fr\alpha4}{\Maxx{x\in A'}{|x(S)|}-\Minn{x\in A'}{|x(S)|}}
\ge\fr\alpha{4n}
.}

As $A'\sbseq A$, the set $A$ itself must contain enough elements, whose projection to $S$ has low Hamming weight:
\m[m_A]{M'\le\sz{A'}
\le\fr{4n}\alpha\tm\ssz[\sz{a(S)}\le\fr{k\tm\sz S}n-\fr\alpha4]{a\in A},}
and the same holds for $B$.

Now fix $B'\sbseq B\sbseq\chs{[n]}k$ and let $q_i\deq\PR[Y\unin B']{Y(i)=1}$ and $\beta\deq\sum_{i=1}^n\sz{q_i-\fr kn}$.
Note that the value of
\m{\sum_{i\in T}\l(\l(\fr kn\r)^2-\PR[(X,Y)\unin A'\times B']{X(i)=Y(i)=1}\r)}
is maximised by
\f{T=\set[{\PR[(X,Y)\unin A'\times B']{X(i)=Y(i)=1}<\l(\fr kn\r)^2}]{i\in[n]},}
so without loss of generality, we fix $T$ to be this set.

For all $i\in T$ it holds that
\mal[P]{\PR{X(i)=Y(i)=1}
&=p_i\tm q_i=(\fr kn+p_i-\fr kn)(\fr kn+q_i-\fr kn)\\
&\ge\l(\fr kn\r)^2-\fr kn\tm\l(\sz{p_i-\fr kn}+\sz{q_i-\fr kn}\r)
-\sz{p_i-\fr kn}\tm\sz{q_i-\fr kn}\\
&\ge\l(\fr kn\r)^2-\fr{2k}n\tm\l(\sz{p_i-\fr kn}+\sz{q_i-\fr kn}\r),}
where the last inequality follows from $i\in T\Then p_i<\fr kn\txt{ or }q_i<\fr kn$.
Accordingly,
\m{\sum_{i\in T}\l(\l(\fr kn\r)^2-\PR[A'\times B']{X(i)=Y(i)=1}\r)
\le\fr{2k}n\tm\sum_{i\in T}\l(\sz{p_i-\fr kn}+\sz{q_i-\fr kn}\r)
\le\fr{2k}n\tm(\alpha+\beta).} %
Therefore, if
\m{\Maxx{\mac{A'\sbseq A\\B'\sbseq B\\\sz{A'},\sz{B'}\ge M'\\T\sbseq[n]}}
{\sum_{i\in T}\l(\l(\fr kn\r)^2-\PR[(X,Y)\unin A'\times B']{X(i)=Y(i)=1}\r)}
\ge\Delta}
then $\alpha\ge\fr{n\Delta}{4k}$ or $\beta\ge\fr{n\Delta}{4k}$.

Let us see what happens if $\alpha$ is non-negligible.
From \bref{m_A}, for some $S\sbseq[n]$:
\m{M'\le\fr{4n}\alpha\tm\ssz[\sz{a(S)}\le\fr{k\tm\sz S}n-\fr\alpha4]{a\in A},}
which can be reformulated as
\m[m_MM]{\fr{M'}M\le\fr{4n}\alpha\tm
\PR[X\unin A]{\sz{X(S)}\le\fr{k\tm\sz S}n-\fr\alpha4}.}

Let $\Cll e_S\deq\left[\sz{X(S)}\le\fr{k\tm\sz S}n-\fr\alpha4\right]$. For a fixed $S$, this event depends only on the value taken by $X$.
First we analyse the probability of $\Cll e_S$ under $X\unin\chs{[n]}k$.
To do that (with accuracy sufficient for our needs), we note that in a sequence of $n$ independent Bernoulli trials with individual success probability $\fr kn$ (next denoted by $\Cl B_{\dr kn}^{\otimes n}$), exactly $k$ successes are observed with probability at least $\fr1n$; moreover, the corresponding conditional distribution is coordinate-wise symmetric.\fn
{
That is, the corresponding marginal distribution is the same at each coordinate.
}
Accordingly,
\m[m_e-nk]{\PR[X\unin\chs{[n]}k]{\Cll e_S}\
\le n\tm\PR[X\sim\Cl B_{\dr kn}^{\otimes n}]{\Cll e_S}
\le n\tm e^{\fr{-n^2\alpha^2}{32k|S|^2}}
\le\exp\l(\ln n-\fr{\alpha^2}{32k}\r)
,}
where the second inequality follows from the Chernoff bound (\clmref{c_Che}), and the last one uses $|S|\leq n$.

Next we claim that the probability of $\Cll e_S$ is unlikely to differ significantly under $X\unin\chs{[n]}k$ and under $X\unin A$ when $A$ is a uniformly-random subset of $\chs{[n]}k$ of size $M$.
Let
\m{\Cll e_S'\deq\left[\PR[X\unin A]{\Cll e_S}\ge\PR[X\unin\chs{[n]}k]{\Cll e_S}+%
\delta\right]
}
for some $\delta<1$ to be fixed later. For a fixed $S$, this event depends only on the content of $A$ (which we now view as a random object).

If instead of choosing $A$ as a subset of size $M$, we would select $M$ times a uniformly-random element of $\chs{[n]}k$ and add it to $A$ (possibly with repetitions), then by the assumption about~$M$, no repetition would occur with probability more than $\dr12$; conditional on that, the process would indeed generate a uniformly-random subset of size $M$.
Let $Y=(Y_i)_{i=1}^M$, where \pl[Y_i] are independent Bernoulli variables that take value ``$1$'' with probability $\PR[X\unin\chs{[n]}k]{\Cll e_S}$, then
\m{\PR[A\unin\chs{\chs{[n]}k}M]{\Cll e_S'}\
\le2\PR{\fr{|Y|}M\ge \PR[X\unin\chs{[n]}k]{\Cll e_S}+\delta}
\le2\tm e^{\fr{-M\delta^2}3}
,}
where the second inequality follows from the Chernoff bound (\clmref{c_Che}).
By the union bound and since $n\ge3$,
\m[m_ep]{\PR[A\unin\chs{\chs{[n]}k}M]{\bigvee_{S}\Cll e_S'}
\le2^{n+1}\tm e^{\fr{-M\delta^2}3}
<\exp\l(n-\fr{M\delta^2}3\r)
.}

Now let $\delta\deq\sq3\tm M^{-\dr13}$.
Combining \bref{m_MM}, \bref{m_e-nk} and \bref{m_ep}, we conclude that if $\alpha\ge\fr{n\Delta}{4k}$, then
\m[m_res]{\fr{M'}M <
\fr{4n}\alpha\tm \l(\exp\l(\ln n-\fr{\alpha^2}{32k}\r) + \delta\r)
\le8n^2\tm e^{-\fr{n^2\Delta^2}{512k^3}} + 14n\tm M^{-\dr13}
}
holds with probability greater than
\f{1-\exp\l(n-M^{\dr13}\r)}
with respect to a uniformly-random $A\sbseq\chs{[n]}k$ of size $M$.
By symmetry, the same is true if $\beta\ge\fr{n\Delta}{4k}$, and therefore true unconditionally.
From \bref{m_res} we conclude that 
\m{\fr{M'}M < 16n^2\tm e^{-\fr{n^2\Delta^2}{512k^3}}}
or
\m{\fr{M'}M < 28n\tm M^{-\dr13}.}
The latter possibility would contradict the lemma assumptions, and the former implies
\m{\Delta^2<\l(\log\l(\fr M{M'}\r)+\log n\r)\tm\fr{2560\tm k^3}{n^2}.}
The result follows.
\prfend

We are ready to implement the lower bound method that has been presented in \sref{s_method}.

\theo[t_II_low]{Let $1\le k\le\fr n3$ and $n^8\le M\le\fr12\tm\chs nk^{\dr12}$.
If $A$ and $B$ are uniformly-random subsets of $\chs{[n]}k$ of size $M$, then
\m{\R_{\dr12}(\IIAB)\ge\R_{\U[A\times B],\dr12}(\IIAB)>
\Min{\fr{\log M}3-8\log n,~\fr{n^2}{93636\tm k^3}-4\log n}
}
holds with probability higher than $1-\exp(n-M^{\dr13}+1)-\PR{\exists a\in A, b\in B~:a\cap b=\emptyset}$.}

Note that the statement of the theorem can be strengthened as follows:
Instead of requiring that $a\cap b\neq\emptyset$ for every possible $a\in A$ and $b\in B$, we could let a uniformly-random pair from $A\times B$ have non-empty intersection with sufficiently high probability $1-\delta$ and allow protocol error strictly higher than $\delta$ (say, looking at $\R_{\dr14+\delta}(\IIAB)$).
Since in this case a valid protocol would be allowed to err whenever $a\cap b=\emptyset$, all the challenges in proving a good lower bound that this work aims to address (as discussed in \sref[s_intro]{s_method}) would still be present.
The reason why we impose the restriction that $a\cap b\neq\emptyset$ for every possible input pair is aesthetic:\ we have been trying to emphasise the \e{syntactic} nature of the guarantee that the intersection was non-empty.

The theorem above can be applied to prove the following.
\crl[c_II_low]{Let $k=n^{\dr35}$ and $M=2^{\dr{\sq[5]n}5}$, then
\m{\R_{\dr12}(\IIAB)\ge\R_{\U[A\times B],\dr12}(\IIAB)
\in\asOm{\sq[5]n}
}
holds with probability $1-2^{-\asOm{\sq[5]n}}$ when $A$ and $B$ are uniformly-random $M$-subsets of $\chs{[n]}k$.
}

The lower bound above is linear in the input size, which is $\log M$.
Accordingly, it is tight and $\R_{\dr12}(\IIAB)\in\asT{\sq[5]n}$ almost always (i.e., for almost all $A$ and $B$). %

\prfstart
Note that
\mal[P]{\PR{\exists a\in A, b\in B~:a\cap b=\emptyset}
&\le M^2\tm\PR[X,Y\unin\chs{[n]}k]{X\cap Y=\emptyset}\\
&=M^2\fr{\chs{n-k}k}{\chs nk}\le M^2\left(\frac{n-k}{n}\right)^k\\
&\le M^2\tm\exp\l(-\fr{k^2}{n}\r)
\le 2^{\dr{-\sq[5]n}{10}}
}
and apply \theoref{t_II_low}.
\prfend[\crlref{c_II_low}]

\prfstart[\theoref{t_II_low}]
Let $\mu$ be the input distribution of $(X,Y)$ -- namely, the uniform distribution over $A\times B$ where $\sz A=\sz B=M$, and assume that $\Cl P$ is a deterministic protocol of complexity $c$ that solves \IIAB\ with error at most $\dr12$ with respect to $\mu$, conditional on $a\cap b\neq\emptyset$ for every $(a,b)\in A\times B$.\fn%
{In the rest of the proof of \theoref{t_II_low}, unless stated otherwise, we implicitly assume the input distribution $\mu$.}
We will keep track of the events
\m{\forall i\in[n]:~\Cll e_i\deq[X(i)=Y(i)=1].}

Informally, we will say that a typical answer ``$i$'' is output by $\Cl P$ with probability at least $\dr1n$ and conditional on the answer ``$i$'', the probability of $\Cll e_i$ is at least $\dr12$.
That is, ``$i$''-labelled rectangles of $\Cl P$ boost the probability of $\Cll e_i$ by roughly $\dr1n$, which must be compensated by lowering the conditional probability of $\Cll e_i$ in the rest of the rectangles of $\Cl P$, and \lemref{l_rec} implies that for that to happen, a typical rectangle must be rather small.

As $\Cl P$ partitions $A\times B$ into $2^c$ rectangles, at least a $(1-n^{-3})$-fraction of the input pairs from $A\times B$ belong to a rectangle with both sides of size at least $M'\deq\fr M{n^3\tm2^c}$.
Denote by $R_+$ the set of all such rectangles, by $R_-$ the rest of \pss[\Cl P] rectangles and let $R=R_+\cup R_-$.
For every $r\in R$, let $\ell(r)$ be the label of the rectangle, i.e., the answer returned by $\Cl P$ when $(X,Y)\in r$.

First, let us show that $\E[(X,Y)\unin A\times B]{\sz{X\cap Y}}$ is unlikely to be too different from $\fr{k^2}n$.
Let $\nu$ be the distribution of the \e{multiset} $B$, resulting from selecting $M$ times uniformly at random an element of $\chs{[n]}k$ and adding it to $B$ (i.e., $|B|\le M$).
Then the inequality below follows from the Chernoff bound (\clmref{c_Che}) and the fact that the distribution of $Y\unin B$ that results from $B\sim\nu$ is uniformly-random:
\m{\forall x_0\in\chs{[n]}k:~
\PR[B\sim\nu]{\E[Y\unin B]{\sz{x_0\cap Y}} > \fr{k^2}n+\fr1{n^2}}
\le e^{-\fr M{n^5}}
.}
On the other hand, the probability that $B\sim\nu$ is a \e{set} is more than
\m{
1-\fr{M^2}{\chs nk}\ge\fr34
,}
in which case $B$ is a uniformly-random subset of $\chs{[n]}k$ of size $M$.
Accordingly,
\m{\PR[|B|=M]{\E[Y\unin B]{\sz{x_0\cap Y}} > \fr{k^2}n+\fr1{n^2}}
<\fr43 \tm e^{-\fr M{n^5}}
<\exp\l(-M^{\dr13}\r)
,}
where $B$ is a uniformly-random subset of $\chs{[n]}k$ of size $M$.
By the union bound,
\mal{\PR[|A|=|B|=M]{\E[(X,Y)\unin A\times B]{\sz{X\cap Y}} > \fr{k^2}n+\fr1{n^2}}
&\le\PR[B]{\exists x_0:\E[Y\unin B]{\sz{x_0\cap Y}} > \fr{k^2}n+\fr1{n^2}}\\
&<\exp\l(n-M^{\dr13}\r)
.}
For the rest of the proof we assume that
$\E{\sz{X\cap Y}}\le\fr{k^2}n+\fr1{n^2}$.

Now we come back to the protocol $\Cl P$.
By the correctness assumption,
\m{\sum_{r\in R}\mu(r)\tm\PRr{\Cll e_{\ell(r)}}{(X,Y)\in r}\ge\fr12.}
On the other hand,
\m{\sum_{r\in R}\mu(r)\tm\sum_{i\in[n]}\PRr{\Cll e_i}{(X,Y)\in r}
=\E{\sz{X\cap Y}}
\le\fr{k^2}n+\fr1{n^2}
.}
Accordingly,
\m{\sum_{r\in R}\mu(r)\tm\sum_{i\neq\ell(r)}\PRr{\Cll e_i}{(X,Y)\in r}
\le\fr{k^2}n+\fr1{n^2}-\fr12
.}

Let $\mu(R_+)\deq\sum_{r\in R_+}\mu(r)$, then $\mu(R_+)\ge1-n^{-3}$ and
\m{\sum_{r\in R_+}\fr{\mu(r)}{\mu(R_+)}
\tm\sum_{i\neq\ell(r)}\PRr{\Cll e_i}{(X,Y)\in r}
&\le\l(\fr{k^2}n+\fr1{n^2}-\fr12\r)\tm\fr1{\mu(R_+)}\\
&\le\l(\fr{k^2}n+\fr1{n^2}-\fr12\r)\tm\l(1+\fr2{n^3}\r)\\
&\le\fr{k^2}n+\fr1{n^2}+\fr{2k^2}{n^4}-\fr12\\
&\le\fr{k^2}n+\fr3{2n^2}-\fr12
.}
Therefore for some $r_0\in R_+$,
\m{\sum_{i\neq\ell(r_0)}\PRr{\Cll e_i}{(X,Y)\in r_0}
\le\fr{k^2}n+\fr3{2n^2}-\fr12
,}
which can be rewritten as
\m{\sum_{i\neq\ell(r_0)}\l(\fr{k^2}{n^2}-\PR[(X,Y)\sim r_0]{\Cll e_i}\r)
\ge\fr12-\fr{3+2k^2}{2n^2}
>\fr16
.}
By \lemref{l_rec}, with probability at least $1-\exp(n-M^{\dr13})$ this implies
\m{\fr{k^{\dr32}}n\tm\sq{\log\l(\fr M{M'}\r)+\log n} >\fr1{306}}
or
\m{\log\l(\fr M{M'}\r) > \fr{\log M}3-5\log n,}
where the former can be rewritten as
\m{\log\l(\fr M{M'}\r) > \fr{n^2}{93636\tm k^3}-\log n.}
The result follows.
\prfend[\theoref{t_II_low}]

\sect*{Acknowledgements}

I am grateful to Pavel Pudl\'ak and Ronald de Wolf for insightful discussions.
I have received many valuable comments from Mika G\"o\"os and several anonymous reviewers.

\end{document}